\documentclass[twocolumn,prl,superscriptaddress,aps,showpacs,nofootinbib]{revtex4-1}

\usepackage{graphicx}
\usepackage{dcolumn}
\usepackage{amsmath}
\usepackage{tikz}
\usepackage{multirow}
\usepackage{bbold}

\newlength{\figwidth}
\newlength{\figwidthb}
\pdfoutput=1

\begin{document}

\newcommand{\hepsilon}{\hat{\epsilon}}
\newcommand\bbone{\ensuremath{\mathbbm{1}}}
\newcommand{\ul}{\underline}
\newcommand{\vl}{v_{_L}}
\newcommand{\vc}{\mathbf}
\newcommand{\be}{\begin{equation}}
\newcommand{\ee}{\end{equation}}
\newcommand{\bk}{{{\bf{k}}}}
\newcommand{\bK}{{{\bf{K}}}}
\newcommand{\cE}{{{\cal E}}}
\newcommand{\bQ}{{{\bf{Q}}}}
\newcommand{\br}{{{\bf{r}}}}
\newcommand{\bg}{{{\bf{g}}}}
\newcommand{\bG}{{{\bf{G}}}}
\newcommand{\hbr}{{\hat{\bf{r}}}}
\newcommand{\bR}{{{\bf{R}}}}
\newcommand{\bq}{{\bf{q}}}
\newcommand{\hx}{{\hat{x}}}
\newcommand{\hy}{{\hat{y}}}
\newcommand{\ha}{{\hat{a}}}
\newcommand{\hb}{{\hat{b}}}
\newcommand{\hc}{{\hat{c}}}
\newcommand{\hd}{{\hat{\delta}}}
\newcommand{\bea}{\begin{eqnarray}}
\newcommand{\eea}{\end{eqnarray}}
\newcommand{\beal}{\begin{align}}
\newcommand{\eeal}{\end{align}}
\newcommand{\ra}{\rangle}
\newcommand{\la}{\langle}
\renewcommand{\tt}{{\tilde{t}}}
\newcommand{\upa}{\uparrow}
\newcommand{\dna}{\downarrow}
\newcommand{\bS}{{\bf S}}
\newcommand{\vS}{\vec{S}}
\newcommand{\dg}{{\dagger}}
\newcommand{\pdg}{{\phantom\dagger}}
\newcommand{\tphi}{{\tilde\phi}}
\newcommand{\cf}{{\cal F}}
\newcommand{\ca}{{\cal A}}
\renewcommand{\ni}{\noindent}
\newcommand{\ct}{{\cal T}}

\title{Spin-orbit coupled systems in the ``atomic'' limit: rhenates, osmates, iridates}
\author{Arun Paramekanti}
\affiliation{Department of Physics, University of Toronto, Toronto, Ontario M5S~1A7, Canada}
\affiliation{Canadian Institute for Advanced Research, Toronto, Ontario, M5G 1Z8, Canada}
\email{arunp@physics.utoronto.ca}
\author{David J. Singh}
\affiliation{Department of Physics and Astronomy, University of Missouri, Columbia, Missouri 65211-7010, USA}
\email{singhdj@missouri.edu}
\author{Bo Yuan}
\affiliation{Department of Physics, University of Toronto, Toronto, Ontario M5S~1A7, Canada}
\author{Diego Casa}
\affiliation{Advanced Photon Source, Argonne National Laboratory, Argonne, Illinois 60439, USA}
\author{Ayman Said}
\affiliation{Advanced Photon Source, Argonne National Laboratory, Argonne, Illinois 60439, USA}
\author{Young-June Kim}
\affiliation{Department of Physics, University of Toronto, Toronto, Ontario M5S~1A7, Canada}
\author{A. D. Christianson}
\affiliation{Materials Science \& Technology Division, Oak Ridge National Laboratory, Oak Ridge, TN-37831, USA}
\affiliation{Neutron Scattering Division, Oak Ridge National Laboratory, Oak Ridge, TN-37831, USA}
\affiliation{Department of Physics \& Astronomy, University of Tennessee, Knoxville, TN-37966, USA}
\date{\today}

\begin{abstract}
%Resonant inelastic X-ray scattering (RIXS) is a versatile tool to probe spin and orbital excitations in solids. 
Motivated by RIXS experiments
on a wide range of complex heavy oxides, including rhenates, osmates, and iridates, we discuss the theory of RIXS
for site-localized $t_{2g}$
orbital systems with strong spin-orbit coupling.
For such systems, we present exact diagonalization results for the spectrum at 
different electron fillings, showing that it accesses ``single-particle'' and ``multi-particle'' excitations. This leads to a simple picture for the energies and 
intensities of the RIXS spectra in Mott insulators such as double perovskites which feature highly localized electrons,
and yields estimates of the spin-orbit coupling and Hund's coupling in correlated
$5d$ oxides. We present new higher resolution RIXS data at the Re-L$_3$ edge in Ba$_2$YReO$_6$ which finds a previously unresolved peak splitting, 
providing further confirmation of our theoretical predictions.
Using {\it ab initio} electronic structure 
calculations on Ba$_2$${\cal M}$ReO$_6$ (with ${\cal M}$=Re, Os, Ir) we show that while the atomic limit yields a reasonable effective Hamiltonian
description of the experimental observations, effects such as $t_{2g}$-$e_g$ interactions and hybridization with oxygen are important.
% in order to 
%connect these effective parameters with a more microscopic description, and to understand variations across the $5d$ oxide series. 
Our
{\it ab initio} estimate for the strength of the intersite exchange coupling shows that, compared to the $d^3$ systems, the exchange
is one or two orders of magnitude weaker in the d$^2$ and d$^4$ materials, which may partly explain the suppression of long-range magnetic order
in the latter compounds. As a way to interpolate between the site-localized picture and our electronic structure band calculations, 
we discuss the spin-orbital levels of the ${\cal M}$O$_6$ cluster. This suggests a possible role for intracluster excitons
in Ba$_2$YIrO$_6$ which may lead to a weak breakdown of the atomic $J_{\rm eff}=0$ picture and to small magnetic moments.
\end{abstract}
\maketitle

\section{Introduction}

In recent years, much attention has been paid to complex oxides of heavy transition elements
%. In these systems, the extended atomic orbitals
%lead to a large octahedral crystal field splitting $\sim 3$eV, typically justifying a projection of the low-energy physics into
%the $t_{2g}$ orbitals. 
where electronic correlations become comparable to the spin-orbit coupling (SOC)
$\lambda$. 
This provides a new route to realizing exotic quantum ground states.\cite{SOCreview}
A large part of this effort has been focussed on the Ir$^{4+}$ iridates with a $5d^5$ configuration, corresponding to a single hole in the $t_{2g}$ orbitals.
\cite{BJKim_PRL2008,Singh2012,Gretarsson2013,Modic2014,Chun2015,bLi2IrO3,gLi2IrO32}
At this filling, the physics is that of a half-filled $j_{\rm eff}=1/2$ band, with the total angular momentum $j_{\rm eff}$ arising from the
coupling of the spin to the effective orbital angular momentum $\ell_{\rm eff}=1$ of the $t_{2g}$ triplet. 
Interest in the spin-orbit coupled materials stems from the possibility of realizing
analogues of high-temperature superconductivity upon electron doping, and exotic magnetic phases such as Kitaev spin liquids and topological semimetals.\cite{Jackeli2009,AIrO3,Valenti_PRB2013,Kimchi2015,Kee_PRL2014,Rachel_PRB2014,YJKim_PRB2014,Perkins_PRX2015,Kee_NComm2015,Valenti_PRB2016,Rachel_PRB2017}
Currently, there is an effort to explore other complex oxides, such as osmates and rhenates, as well as iridates with different valence states,
which may lead to further exotic phenomena at different electron fillings.\cite{GChen2010,GChen2011,Khaliullin2013,Akbari2014,Trivedi2015,Chaloupka2016,Trivedi2017,Fiete_2017,Gong2018}
An important step in this programme is to elucidate the `atomic'  interactions which govern the
local physics, which then feeds into understanding how such local degrees of freedom interact and organize at longer length scales. Here, we discuss this
step in the context of double perovskite materials using a theoretical analysis of resonant inelastic X-ray scattering (RIXS), exact diagonalization studies of the
single-site problem with SOC at different electron fillings (d$^2$, d$^3$, d$^4$), and complementary {\it ab initio} electronic structure calculations.
\footnote{This manuscript has been authored by UT-Battelle, LLC under Contract No. DE-AC05-00OR22725 with the U.S. Department of Energy.  The United States Government retains and the publisher, by accepting the article for publication, acknowledges that the United States Government retains a non-exclusive, paid-up, irrevocable, world-wide license to publish or reproduce the published form of this manuscript, or allow others to do so, for United States Government purposes.  The Department of Energy will provide public access to these results of federally sponsored research in accordance with the DOE Public Access Plan (http://energy.gov/downloads/doe-public-access-plan).}

RIXS has proven to be a particularly valuable tool to explore spin and orbital excitations,
and there has been extensive experimental \cite{YJKim_PRL2002,Gog_PRL2005,YJKim_PRB2010,BJKim_PRL2012,Hill_PRL2013,YJKim_PRB2013,Sala2014PRB,Calder2016}
and theoretical work \cite{Brink_JPCS2005,BrinkEPL2006,Ament_PRB2007,Brink_EPL2007,Forte_PRB2008,Brink_PRL2009,Ament2011,Veenendaal2011,
Brink_PRL2012,Brink_PRL2013,Sala2014,Halasz2016,Demler_PRB2016,Klich_PRB2016,Klich_PRB2017} in this area
(see Ref.~\onlinecite{RIXSReview_RMP2011} for a review).
In this paper, we discuss the theory of RIXS for systems with highly localized $t_{2g}$ electronic states at various fillings.
Using exact diagonalization (ED) calculations of the RIXS spectrum, we show that we can quantitatively extract the spin-orbit and Hund's couplings,
and explain both the energies and the spectral intensities 
observed in experiments on rhenates, osmates, and iridates.
We also present new experimental high resolution RIXS results on Ba$_2$YReO$_6$ at the Re $L_3$ edge
which finds a peak splitting in the spectrum, in perfect agreement with our theoretical predictions. This splitting was not resolved in previous
experiments at the Re $L_2$ edge.
This paper thus extends and generalizes previous well-known
work on RIXS for $5d^5$ iridates,\cite{Ament2011} and provides a useful companion to a recent study of the RIXS operator
in $t_{2g}$ spin-orbital systems.\cite{Khaliullin_PRB2017}

Additionally, in order to complement this effective Hamiltonian study, we have carried out {\it ab initio} electronic structure calculations for the 
cubic double perovskites Ba$_2$YReO$_6$, Ba$_2$YOsO$_6$, and Ba$_2$YIrO$_6$. 
This permits us to understand material-to-material variations of these effective parameters across the $5d$ oxides, 
and to show that $t_{2g}$-$e_g$ interactions and hybridization with the ligand ions (oxygen) play a key role when we attempt to 
connect the parameters of the effective Hamiltonian with a more microscopic theory. 
Furthermore, our
{\it ab initio} estimates for the strength of the magnetic exchange coupling between moments in these materials shows that, compared to the osmates, the exchange
is one or two orders of magnitude weaker in the rhenates and iridates. This could explain the robust magnetic long range order observed in the
osmates, which should be contrasted with weak ordering tendencies in the latter compounds.

Based on our {\it ab initio} calculations, hybridization of the transition metal ion with the surrounding oxygen octahedral cage plays an
important role in complex $5d$ oxides. This leads us to examine the spin-orbital states on the ${\cal M}$O$_6$ metal-oxygen octahedra, which could be useful
in future studies of the effect of extended interactions on such clusters as a way to bridge the gap between ED and DFT results on Ba$_2$YIrO$_6$. 
%A schematic picture of the 
%energy levels on the cluster suggests possible intra-cluster exciton
%mechanisms in Ba$_2$YIrO$_6$ which could lead to a weak breakdown of the strictly atomic $J_{\rm eff}=0$ picture and generate small magnetic moments.

\section{RIXS for highly localized states}

The Kramers-Heisenberg expression \cite{Brink_JPCS2005,BrinkEPL2006,Ament_PRB2007,Brink_EPL2007,Forte_PRB2008,RIXSReview_RMP2011} 
for the two-photon RIXS scattering cross section is given by
\bea
\frac{d^2\sigma}{d\Omega dE_{\rm in}} &=& \frac{E_{\rm out}}{E_{\rm in}} \sum_f
\left| \sum_n \frac{\la f| T^\dagger | n \ra \la n | T | g \ra}{E_g - E_n + E_{\rm in} + i \frac{\Gamma_n}{2}} \right|^2 \nonumber \\
&\times& \delta(E_g - E_f + E_{\rm in}-E_{\rm out}).
\eea
Here, $g,n,f$ refer to ground (initial) state, intermediate state, and final state, respectively with corresponding energies $E_g, E_n, E_f$, and $\Gamma_n$ is the inverse lifetime
of the intermediate state.
$E_{\rm in}=\hbar\omega_{\rm in}$ and $E_{\rm out}=\hbar\omega_{\rm out}$ are the incoming and outgoing photon energies, and the $\delta$-function enforces energy conservation.
Within the dipole approximation for the photon field,
the transition is induced by the dipole operator $T \sim \hat{\epsilon} \cdot \br$,
where $\hat{\epsilon}$ denotes the photon polarization, which we label as $\hat{\epsilon}_{\rm in}$ for the incoming photon which
excites from the ground state (which enters in the $\la n | T | g \ra$ matrix element above), and $\hat{\epsilon}_{\rm out}$ for the outgoing photon which
de-excites into the final state (which enters in the $\la f| T^\dagger | n \ra$  matrix element above).
On resonance, with $E_{\rm out} \approx E_{\rm in}$ (since the energy transfer $E_f-E_g \ll E_{\rm in}, E_{\rm out}$), the cross section simplifies to
\bea
\frac{d^2\sigma}{d\Omega dE_{\rm in}} &\approx& \frac{1}{A} \sum_f
\left| \sum_n \la f| T^\dagger | n \ra \la n | T | g \ra \right|^2 \nonumber \\
&\times& \delta(E_g \!-\! E_f \!+\! E_{\rm in} - E_{\rm out})
\label{eq:simplify}
\eea
where the prefactor $A = \left|E_g - \bar{E}_n + E_{\rm in} + i \frac{\bar{\Gamma}_n}{2} \right|^2$,
with $\bar{E}_n, \bar{\Gamma}_n$ being the {\it average} energy and inverse lifetime of the intermediate states.
This approximation, which is valid for short core-hole lifetime,\cite{BrinkEPL2006,Ament_PRB2007}
allows us to ignore intermediate state interactions between the core-hole and other electrons.
We show below that the resulting spectra are in good agreement with experiments, providing a phenomenological
justification for this approximation.

\begin{figure*}[tb]
	\begin{center}
\includegraphics[width=0.5\textwidth]{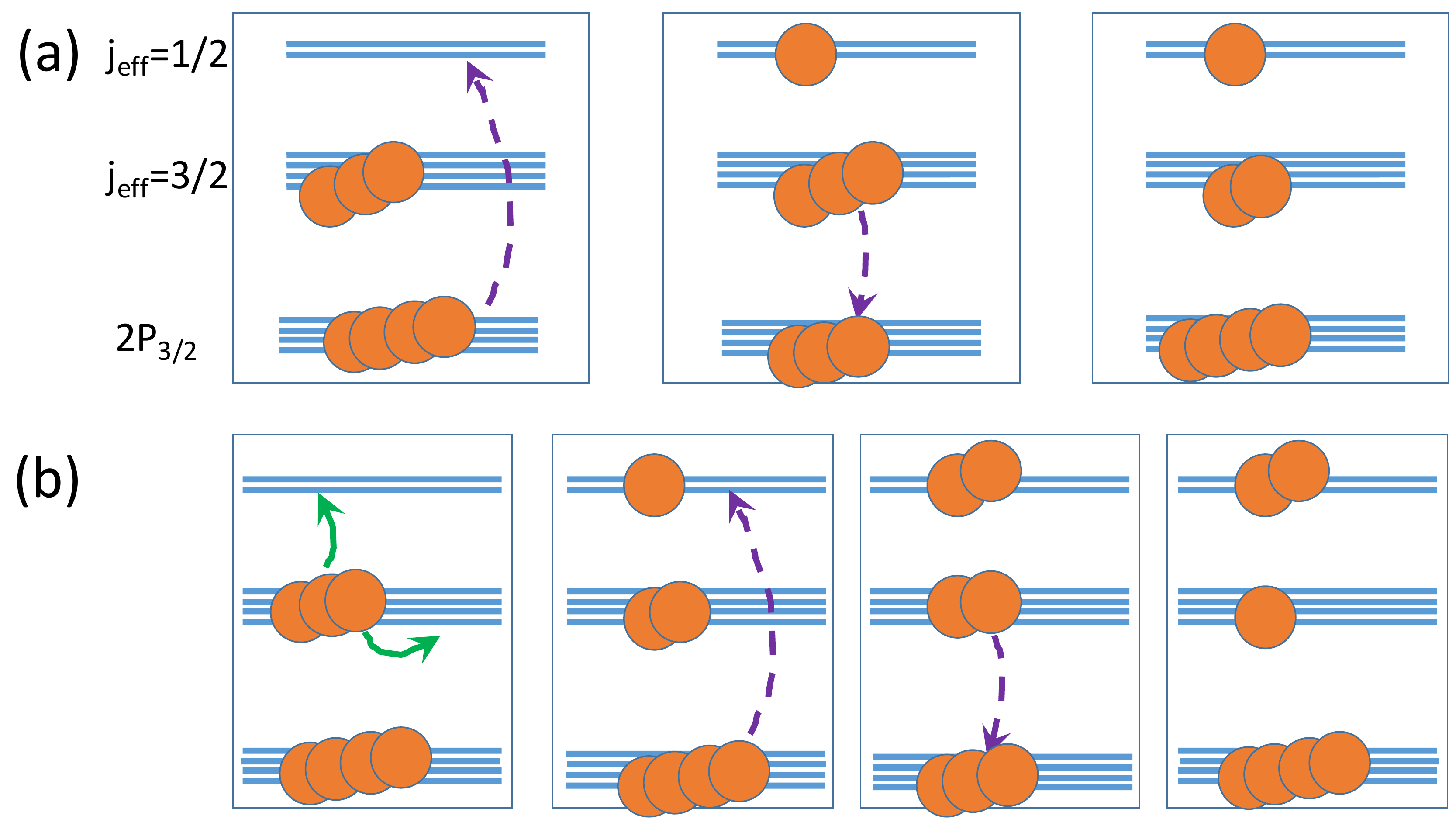}\includegraphics[width=0.5\textwidth]{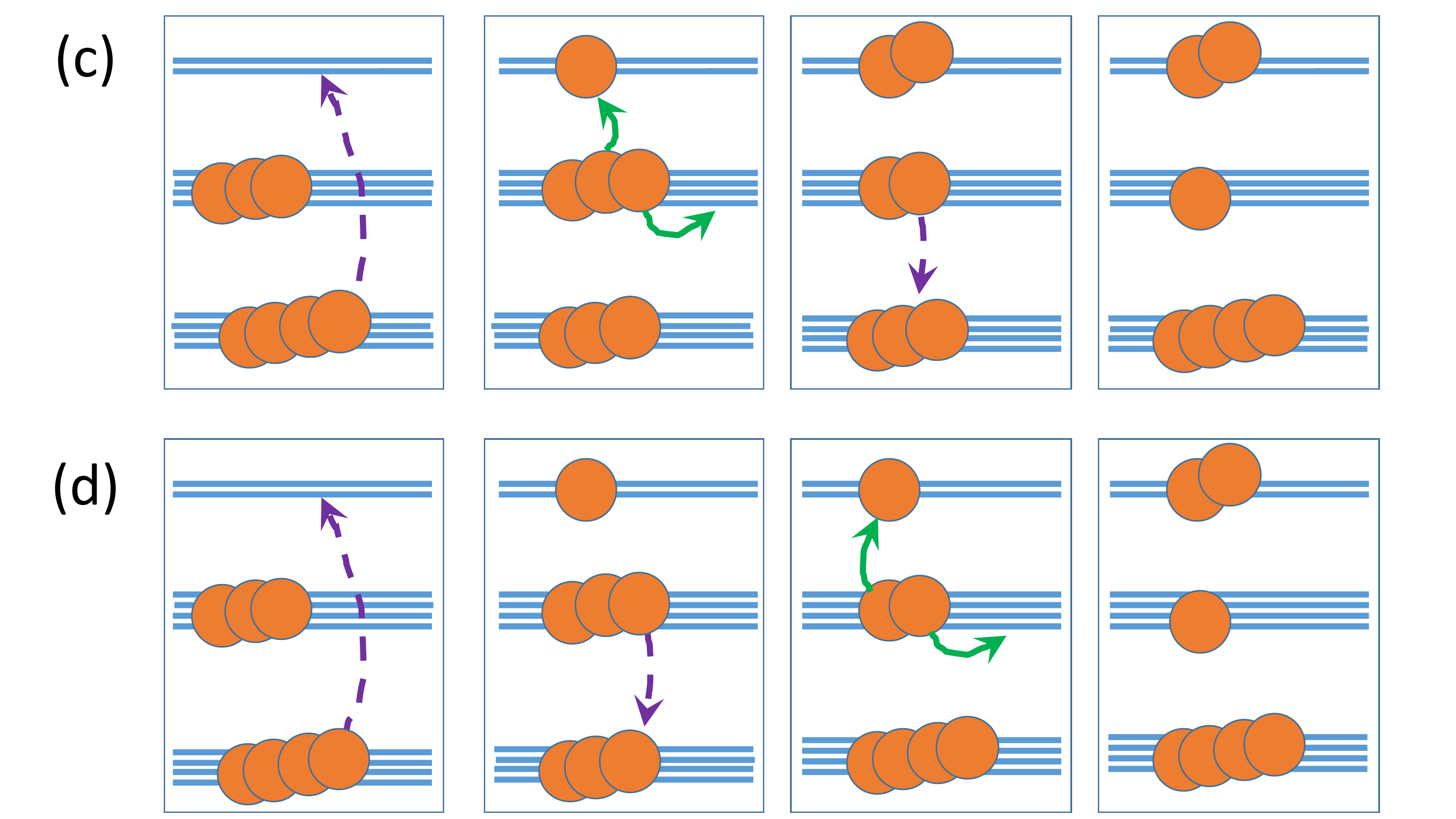}
		\caption{(Color online) Schematic picture of the $L_3$ edge inelastic RIXS processes (for the 5d$^3$ osmates), with dashed (purple) lines indicating photon-induced transitions
		and solid (green) lines indicating interaction-induced transitions. 
		(a): Noninteracting case, where the two-photon process consists of
		a core electron getting excited into the $j_{\rm eff}=1/2$ manifold followed by a de-excitation transition from $j_{\rm eff}=3/2$ back into the core level. This
		would lead to a RIXS peak at $3\lambda/2$.
		(b),(c),(d): Interacting case, where the local Hund's coupling leads to visible multi-particle
		excitations. Within perturbation theory, interactions can scatter electrons into higher energy states as depicted by the solid (green) lines for the (b) initial, (c) intermediate, and
		(d) final states. For $J_H < \lambda$, this would lead to $2$-particle peaks near $3\lambda$ but with suppressed intensity $\sim (J_H/\lambda)^2$.}		
		\label{fig:schematic}
	\end{center}
\end{figure*}

RIXS excites an electron from a highly spin-orbit coupled core level into the
relevant $d$-orbitals; here, we focus on excitation into the $t_{2g}$ states. This leads to an intermediate state with a core-hole and an added electron in the $t_{2g}$ orbitals.
These intermediate states decay on the timescale of the core-hole lifetime $\sim 1/\Gamma_n$, leaving the original $t_{2g}$ electrons in a final excited spin-orbital state.
We can thus consider simplified transition matrix elements \cite{Trivedi2015}
\bea
\la n | T | g \ra &=& \hepsilon^\alpha_{\rm in} \la n | p^\dg_{\beta\sigma}  d^\dg_{\alpha\beta\sigma}  | g \ra\\
\la f | T^\dagger | n \ra &=& \hepsilon^\mu_{\rm out} \la f | d^\pdg_{\mu\nu\sigma'} p^\pdg_{\nu\sigma'} | n \ra.
\eea
Here $p^\dg_{\alpha\sigma}$ creates a $2P$ core-hole in orbital
$\alpha$ ($\alpha=p_x,p_y,p_z$) with spin $\sigma$, while $d^\dg_{\alpha\beta\sigma}$ creates a $d$-electron in the $t_{2g}$ orbital ($d_{yz},d_{zx},d_{xy}$)
with spin $\sigma$, and we have restricted attention to parity-allowed nonzero dipole matrix elements. 
Using this, we arrive at the following expression for the RIXS cross-section:
\bea
\frac{d^2\sigma}{d\Omega dE_{\rm in}} &\propto&  \! \sum_f
\left| \sum_n \hepsilon^\mu_{\rm out} \hepsilon^\alpha_{\rm in}  \la f | d^\pdg_{\mu\nu\sigma'} p^\pdg_{\nu\sigma'} | n \ra 
\la n | p^\dg_{\beta\sigma}  d^\dg_{\alpha\beta\sigma}  | g \ra
 \right|^2 \nonumber \\
 &\times&
\delta(E_g \!-\! E_f \!+\! E_{\rm in} - E_{\rm out}).
\eea

The core-level part of the process consists of exciting a single core-hole from the core-vacuum and de-exciting back into 
the vacuum, with intermediate states for the core-hole being $2P_{1/2}$ ($L_2$ edge) or $2P_{3/2}$ ($L_3$ edge). Let us denote
the corresponding $P$-matrix elements as $M^{J}_{\nu\sigma';\beta\sigma}$ with $J=1/2,3/2$.
%(see Appendix). 
This
leads to $(d^2\sigma/d\Omega dE_{\rm in}) \propto  {\cal I}_J(\omega)$, where
\bea
{\cal I}_J(\omega) &\equiv& \!\! \sum_f
\left| \sum_n \hepsilon^\mu_{\rm out} \hepsilon^\alpha_{\rm in} M^J_{\nu\sigma';\beta\sigma} \la f | d^\pdg_{\mu\nu\sigma'} | n \ra 
\la n | d^\dg_{\alpha\beta\sigma}  | g \ra
 \right|^2 \nonumber \\
 &\times&
\delta(E_g \!-\! E_f \!+\! \hbar\omega),
\label{eq:rixs}
\eea
with $\hbar\omega = E_{\rm in} - E_{\rm out}$ being the photon energy loss.
We continue to use the notation $g,n,f$, for ground, intermediate, and final states, but henceforth these will refer to only the $t_{2g}$ states. 

The matrix $M^J$ for the core hole for the $L_2$ and $L_3$ edges is given by
\bea
M^{J=1/2} &=& \frac{1}{3} (1-\vec L \cdot \vec S) \\
M^{J=3/2} &=& \frac{1}{3} (2+\vec L \cdot \vec S)
\eea
where $L,S$ refer to the oxygen $2P$ hole orbital- and spin angular momentum operators.
Labelling the $P$-states as $(\mu\sigma) = (p_x\!\!\upa,p_y\!\!\upa,p_z\!\!\upa,p_x\!\!\dna,p_y\!\!\dna,p_z\!\!\dna)$, we can
explicitly write out
\bea
M^{J=1/2} = \frac{1}{3} 
\begin{pmatrix}
1 & -i & 0 & 0 & 0 & -1 \\
i & 1 & 0 & 0 & 0 & -i \\
0 & 0 & 1 & 1 & i & 0 \\
0 & 0 & 1 & 1 & i & 0 \\
0 & 0 & -i & -i & 1 & 0 \\
-1 & i & 0 & 0 & 0 & 1
\end{pmatrix}
\eea
and $M^{J=3/2}=\mathbb{1}-M^{J=1/2}$.
The sum over all intermediate $d$-states in Eq.~\ref{eq:rixs}
can be done, which leads to
\bea
{\cal I}_J(\omega) &\equiv& \!\! \sum_f
\left| \hepsilon^\mu_{\rm out} \hepsilon^\alpha_{\rm in} M^J_{\nu\sigma';\beta\sigma} \la f | d^\pdg_{\mu\nu\sigma'} d^\dg_{\alpha\beta\sigma}  | g \ra
 \right|^2 \nonumber \\
&\times& \delta(E_g \!-\! E_f \!+\! \hbar\omega).
\label{eq:rixs2}
\eea
Below, we will discuss a physical picture for the excitations, before turning to exact diagonalization results.

Note that everywhere below, we will work with the single particle spin and orbital basis states for the $t_{2g}$ electrons. However,
 when interactions are absent or weak compared with SOC, we will refer to the $j_{\rm eff}=1/2,3/2$ eigenstates. Furthermore,
 the core hole state is treated separately, ignoring its intermediate state interactions which is justified in the short-lifetime limit.
%, and we can simplify the above
%expression to its final form:
%\bea
%{\cal I}(\omega) &\equiv& \!\! \sum_f
%\left| \hepsilon^\alpha_{\rm in}  \hepsilon^\mu_{\rm out} \varepsilon^\pdg_{\alpha\mu\nu} \la f | L^\nu  | g \ra
% \right|^2
%\delta(E_g \!-\! E_f \!+\! \hbar\omega)
%\label{eq:rixs3}
%\eea
%where $L^\nu = - i \varepsilon^\pdg_{\nu\alpha\beta} \sum_\sigma d^\dg_{\alpha\sigma} d^\pdg_{\beta\sigma}$ is the orbital angular momentum operator
%counting both spins $\sigma$. RIXS in this setting thus directly probes the spectrum of orbital angular momentum fluctuations. In systems with spin-orbit coupling, 
%it thus also probes aspects of spin fluctuations.

\section{Physical picture of excitations}

In the absence of electron-electron interactions, the RIXS process is schematically illustrated in Fig.~\ref{fig:schematic}(a). Here, we depict spin-orbit split $t_{2g}$ levels,
having an effective (single-particle) angular momentum states, with a low energy $j_{\rm eff}=3/2$ quartet and a higher energy $j_{\rm eff}=1/2$ doublet.
These are split by $3\lambda/2$ where $\lambda$ is the spin-orbit coupling. We consider
a filling corresponding to a 5d$^3$ configuration (e.g., osmates), and depict the photon-induced transitions by dashed (purple) lines.
For the $L_3$ edge, the incoming photon excites a core electron from 
$2P_{3/2}$ into the higher energy
$j_{\rm eff}=1/2$ state, followed by a de-excitation from the lower energy $j_{\rm eff}=3/2$ manifold into the core-level. Such inelastic processes would lead to only
a {\it single} peak at $\hbar\omega=3\lambda/2$.

Next, let us consider interactions between electrons in the $t_{2g}$ manifold, given by
\bea 
\!\!\!\! H_{\rm int} \! &=&\! \frac{U}{2} :(\sum_{\ell} n^\pdg_{\ell})^2:
-5 \frac{J_H}{2} \sum_{\ell<\ell'} n^\pdg_{\ell}
n^\pdg_{\ell'} \nonumber \\
\! &-&\! 2 J_H \sum_{\ell< \ell'} \vec S^\pdg_{\ell}
\cdot \vec S^\pdg_{\ell'} + J_H \sum_{\ell \neq \ell'} d^\dg_{\ell\upa}
d^\dg_{\ell\dna} d^\pdg_{\ell' \dna} d^\pdg_{\ell' \upa}
\label{Hint}
\eea
with $::$ denoting normal ordering. Here the various terms in the Kanamori interaction are: (i) the total ``charging energy'' to change the electron number at a site, 
(ii) the difference term between interorbital and intraorbital charge repulsion, (iii) the Hund's exchange between spins in different orbitals, and (iv) singlet pair hopping
between orbitals. While
the first term is governed by the ``Hubbard $U$'', the latter three interactions are all set by the Hund's coupling.
Since RIXS is a number conserving process, and the intermediate state of the $t_{2g}$ orbitals
plays no role in the expression in Eq.~\ref{eq:rixs2}, the charge repulsion $U$
plays no role in determining ${\cal I}_J(\omega)$. The interactions relevant to RIXS are therefore parameterized by a 
single energy: the Hund's coupling $J_H$.

Such interaction effects will lead to {\it multiple} peaks in the RIXS spectrum, deviating from the single-particle expectation.
At the perturbative level, this stems from two reasons. First, many-body effects will split the degeneracies associated with the single-particle states; this will split the peak at $3\lambda/2$ into multiple peaks separated by the interaction energy scale $J_H$. Second, 
interactions can perturbatively excite electrons 
into higher energy single-particle states. This is shown in Figs.~\ref{fig:schematic}(b-d), where interactions excite electrons between 
$j_{\rm eff}=3/2$ and $j_{\rm eff}=1/2$ states as shown by the solid (green) line. This can happen in the ground, 
intermediate, or final states, and it leads to transitions into final states with {\it two} electrons excited across the spin-orbit gap. Such `multi-particle' excitations will produce a 
second set of peaks around an energy $\sim 3\lambda$. For small $J_H/\lambda$, these secondary peaks will have an intensity $\sim (J_H/\lambda)^2$. This is in addition to any suppression of matrix elements arising from
quantum numbers (i.e., selection rules).

Below, we will use the expression in Eq.~\ref{eq:rixs2}, and present results from a (non-perturbative)
numerical computation using exact diagonalization for the $g,n,f$ states of the
$t_{2g}$ orbitals. While we have presented preliminary results for the case of the iridates in previous work, we focus here on other fillings, which are 
also relevant to the osmates and rhenates.

\section{Exact diagonalization results}

\subsection{Mode energies}

We have used the $t_{2g}$ orbital basis with SOC and the Kanamori interaction, and numerically computed the eigenstates and the RIXS intensity from
Eq.~\ref{eq:rixs2} using exact diagonalization. 
The projection to the $t_{2g}$ levels is justified by the large crystal field splitting as seen in the ab initio results (discussed below).
We consider different fillings $d^2,d^3,d^4$, and compare the resulting energies in the spectrum to the 
experimental
results \cite{Taylor_PRL2017, Yuan_PRB2017,Nag2017,Kusch2018} for various $5d$ Mott insulating oxides: (i) Ba$_2$YReO$_6$ ($d^2$ rhenate), (ii) Ca$_3$LiOsO$_6$ and Ba$_2$YOsO$_6$ ($d^3$ osmates), and
(iii) Sr$_2$YIrO$_6$, Sr$_2$GdIrO$_6$, and Ba$_2$YIrO$_6$ ($d^4$ iridates). A best fit of the excitation energies to previously published 
experimental spectra allows us to extract the SOC strength 
$\lambda$ and the Hund's coupling $J_H$. The results are summarized in Table I, where we also show the excitation energies from theory and experiments.  
The agreement is good, showing that the projection to the $t_{2g}$ sector yields an effective description of the RIXS data. 

\begin {table}[tbh]
%\begin{center}
    \begin{tabular}{| c | c | c | c | c | c| c| c |}
    \hline
    Material & & $\lambda$ & $J_H$ & Peak 1 & Peak 2 & Peak 3 & Peak 4 \\ \hline
    \multirow{2}{*}{} {Ba$_2$YReO$_6$} & Ex & & & [{\it 0.40}] &  [{\it 0.50}]  & {\it 0.83} & {\it 1.85} \\ (ref.\onlinecite{Yuan_PRB2017},this) & Th & 0.380  &  0.260 & 0.41 & 0.47 & 0.89 & 1.83 \\ \hline
     \multirow{2}{*}{} {Ba$_2$YOsO$_6$} & Ex & & & {\it 0.745} & {\it 0.971} & {\it 1.447} & {\it 1.68} \\ (ref.\onlinecite{Taylor_PRL2017}) & Th & 0.335  & 0.275   & 0.75 & 0.91 & 1.46 & 1.71 \\ \hline
    \multirow{2}{*}{} {Sr$_2$YIrO$_6$}  & Ex & & & {\it 0.39} & {\it 0.66} & {\it 1.30} & {\it $\sim$2} \\ (ref.\onlinecite{Yuan_PRB2017}) & Th & 0.425  &  0.250 & 0.41 & 0.64 & 1.31 & 2.06 \\ \hline
     \multirow{2}{*}{} {Ba$_2$YIrO$_6$}  & Ex & & & {\it 0.35} & {\it 0.60} & {\it 1.18} & {\it - } \\ (ref.\onlinecite{Nag2017}) & Th & 0.385  &  0.230 & 0.37 & 0.58 & 1.19 & 1.88 \\
    \hline
    \end{tabular}
    \caption{Table showing the optimal $\lambda, J_H$ (in eV) in different materials deduced from fitting the theoretical calculations to experimental RIXS excitation energies.
We present the comparison of the observed (from Refs.~\onlinecite{Taylor_PRL2017, Yuan_PRB2017,Nag2017})
RIXS peak energies (in eV) (top, `Ex', in italics) with the corresponding theoretical values (bottom, `Th') for the optimal parameter set computed here. For Ba$_2$YReO$_6$,
peaks $\#1$ and $\#2$ (in square brackets) were not previously resolved with $L_2$ edge RIXS, but are resolved here using the $L_3$ edge (see text and Fig.~\ref{newdata}).
For Sr$_2$YIrO$_6$, peak $\#4$
is seen as a very weak $\sim 2$eV feature, and it is absent in the nominally cubic Ba$_2$YIrO$_6$. Results for Ca$_3$LiOsO$_6$ (Ref. \onlinecite{Taylor_PRL2017})
and Sr$_2$GdIrO$_6$ (Ref.~\onlinecite{Yuan_PRB2017}) are nearly identical to Ba$_2$YOsO$_6$ and Sr$_2$YIrO$_6$ respectively.}
%\end{center}
\end{table}

\begin{figure*}[tbh]
	\begin{center}
		\includegraphics[width=0.33\textwidth]{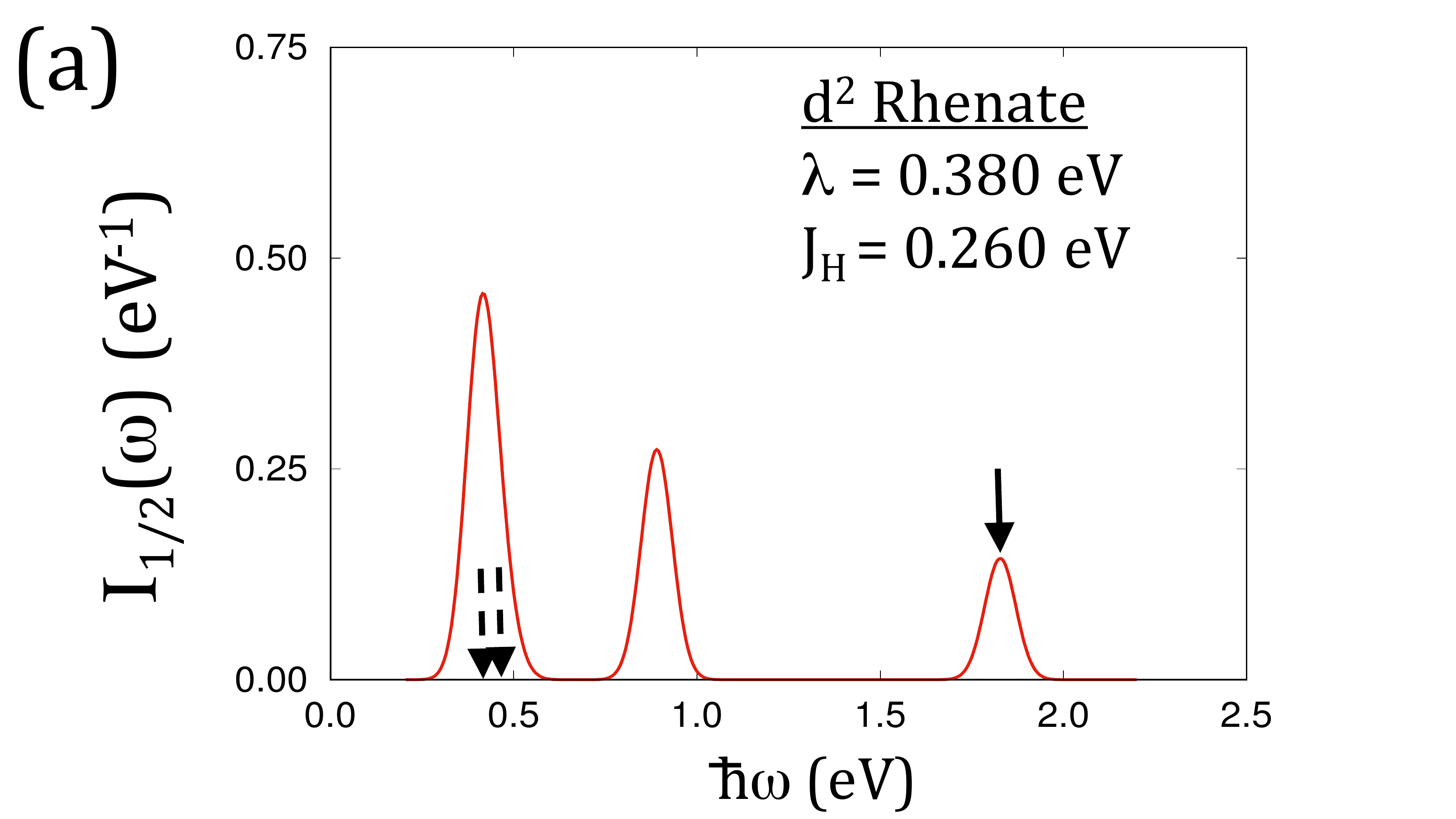}\includegraphics[width=0.33\textwidth]{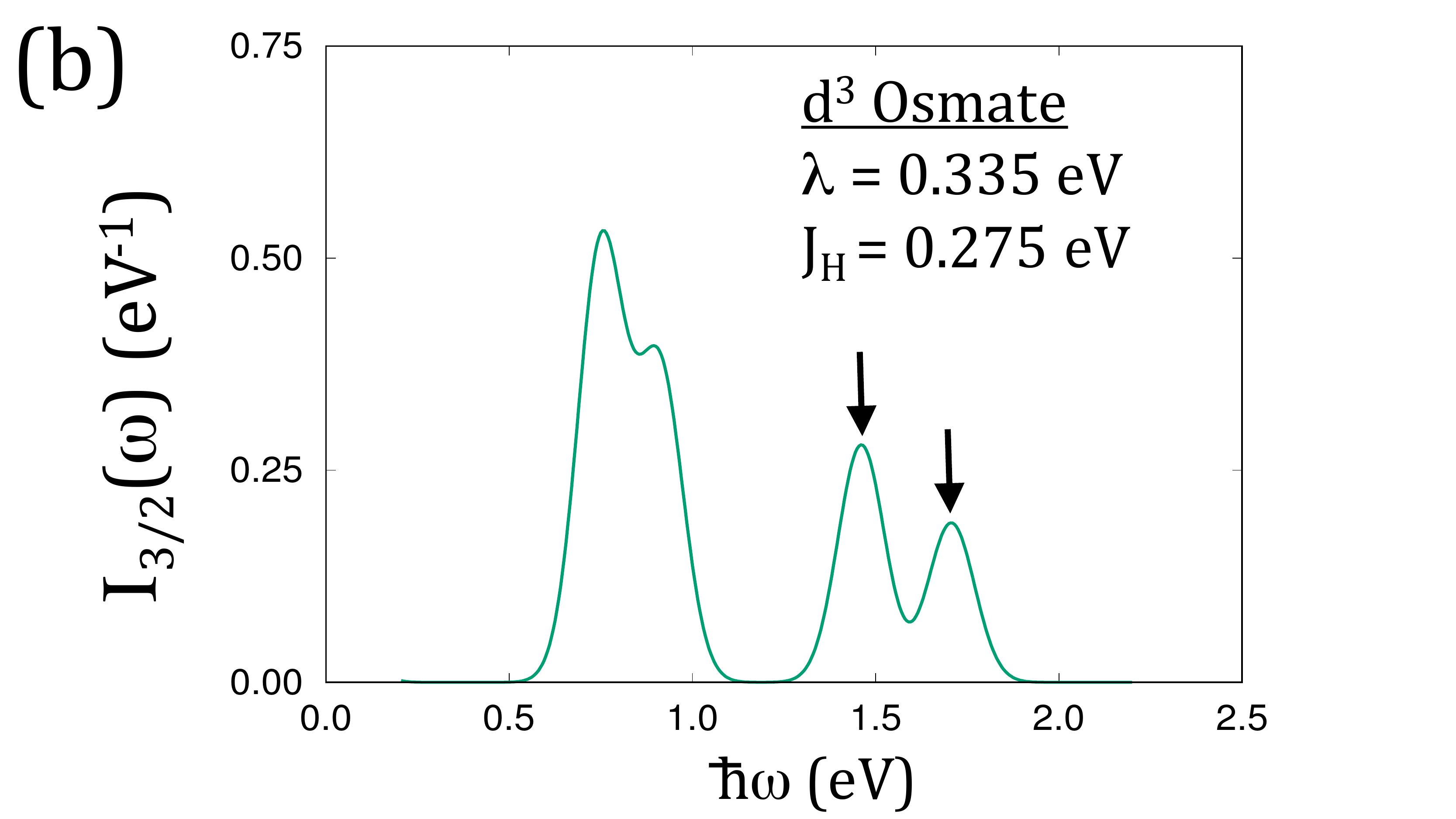}\includegraphics[width=0.33\textwidth]{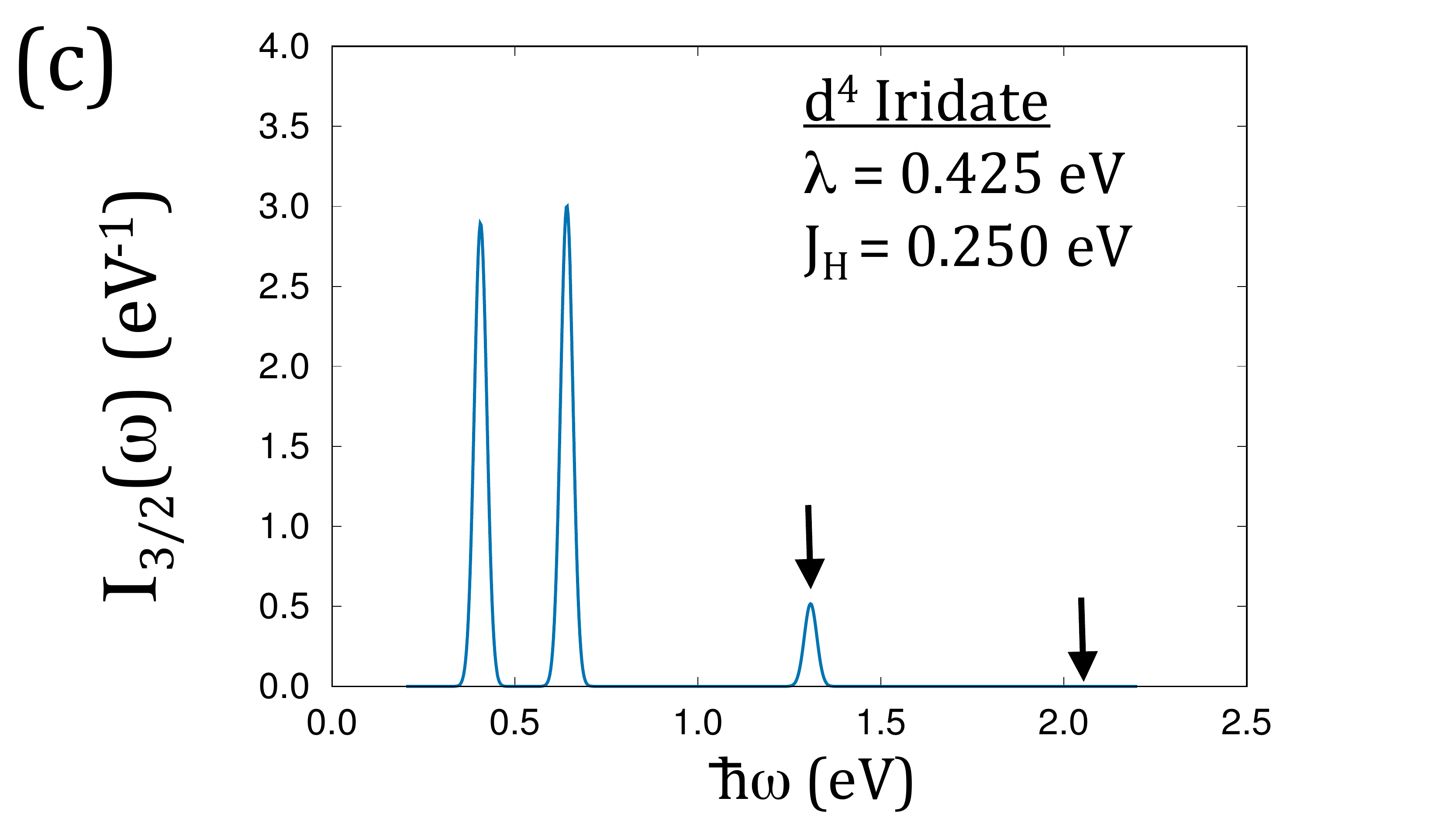}
		\caption{(Color online) Theoretically computed RIXS spectrum for (a) $d^2$ rhenates ($L_2$ edge), (b) $d^3$ osmates ($L_3$ edge), and (c) $d^4$ iridates 
		($L_3$ edge) broadened by instrumental resolution. We have used shown best fit values for $\lambda$ (SOC) and $J_H$ (Hund's coupling).
		In (a), dashed arrows indicate two closely spaced peaks, which are not resolved in $L_2$ edge RIXS;
		see Fig.~\ref{newdata} for new $L_3$ edge RIXS which detects this splitting.
		In all panels, solid arrows indicate `two-particle' transitions due to Hund's coupling.}
		\label{fig:spectra}
	\end{center}
\end{figure*}

\subsection{Spectral intensities}

RIXS experiments are typically carried out in the `horizontal geometry' where the in-photon polarization lies in the scattering plane, with 
the scattering angle to be $2\theta=90^\circ$, so that $\hepsilon_{\rm in} \cdot \hepsilon_{\rm out} =0$. 
Fig.~\ref{fig:spectra} shows the spectrum computed for this scattering geometry. Here, we average over $\hepsilon_{\rm out}$ 
with $\hepsilon_{\rm in} \cdot \hepsilon_{\rm out} =0$. The precise incident polarization direction does not 
matter since the results are rotationally invariant (so single crystals and powder samples should yield the same result in this
`atomic limit').
We have chosen experimentally relevant values for the resolution, with a full width at half 
maximum (FWHM) of $100$meV (rhenates, $L_2$ edge), $150$meV (osmates, $L_3$ edge), and $40$meV (iridates, $L_3$ edge). In all cases, 
the two lower energy peaks (energies $\lesssim 1$eV), which, as discussed above,
arise from single-particle excitations across the spin-orbit gap have higher spectral weight, while the higher energy peaks which are due to multiparticle
excitations have weaker intensity. This is in reasonable agreement with experiments across all materials. The iridates, which have a non-degenerate
$J_{\rm eff}=0$ ground state are most robust to interaction effects, and exhibit negligible intensity for two-particle excitations.

\subsection{$L_3$ edge RIXS for rhenates}

RIXS measurements at Re L3 edge (E$_i$=10.537~keV) were carried out at the 27ID-B beam line at Advanced Photon Source. The same polycrystalline sample of Ba$_2$YReO$_6$ used in Ref. 51 was used. The beam was monochromatized by Si(111) double-crystal and a Si(119) channel-cut secondary crystal. A spherically diced Si(119) analyser with 2m radius of curvature was used to achieve an overall energy resolution of 60meV.  A horizontal scattering geometry with scattering angle $2\theta=90^{\circ}$ was used to minimize elastic background. The measurement was carried out at room temperature.
This measurement allows us to resolve the splitting between two low energy peaks at $\hbar\omega=0.40$eV and $0.50$eV, which 
was unresolved in previous RIXS measurements and appeared as a single peak. This data provides further quantitative confirmation of 
our theoretical predictions, and it is included in Table I. Note that the intensity of the highest energy $1.8$eV peak, seen clearly in
the published L2 edge data \cite{Yuan_PRB2017}, vanishes in the L3 edge data; since the peak positions themselves do not depend on which
edge is used in the RIXS, we have chosen
to make the theory plot in Fig.~\ref{fig:spectra} for the L2 edge for which the $1.8$eV peak is also clearly visible.

\subsection{Discussion}

Our model Hamiltonian in the $t_{2g}$ spin-orbital sector provides a good description of the RIXS data, with comparable strengths
of the SOC $\lambda$ and Hund's coupling $J_H$. Thus, the site-localized limit provides a good starting point to understand these double perovskites.
However, a strict projection of the physics to
$t_{2g}$ orbitals completely ignores the $e_g$ states. Furthermore, the $d$-orbitals 
of the transition metal ions are expected to hybridize with the neighboring oxygens. Such effects could be important in relating
the $t_{2g}$ model Hamiltonian parameters to a more microscopic description.

For instance, Table I shows a small, but systematic, difference between the RIXS peak energies in polycrystalline cubic Ba$_2$YIrO$_6$ from
Ref.~\onlinecite{Nag2017}, and those reported on single crystals of Sr$_2$YIrO$_6$ and Sr$_2$GdIrO$_6$ in Ref.~\onlinecite{Yuan_PRB2017}. This
must be attributed to the different size of Ba ion compared with Sr, which leads to slight differences in bond lengths and angles of 
the IrO$_6$ octahedra, suggesting that Ir-O hybridization might lead to
renormalization of $\lambda$ and $J_H$.
Table I also shows that the inferred  SOC $\lambda$ for Os is smaller than both Re and Ir, while the corresponding Hund's coupling is 
slightly larger. Again, such a non-monotonic trend across the $5d$ series reflects how the two microscopic effects discussed above might renormalize 
the parameters of the effective Hamiltonian. Such effects may be phenomenologically accounted for by going beyond
the Kanamori Hamiltonian, for instance by modifying the coupling strengths appearing in Eq.~\ref{Hint} as done in 
Ref.~\onlinecite{Taylor_PRL2017} for the osmates. Below, we
use electronic structure calculations to provide an {\it ab initio} perspective.

\begin{figure}[tbh]
\begin{center}
\includegraphics[width=0.5\textwidth]{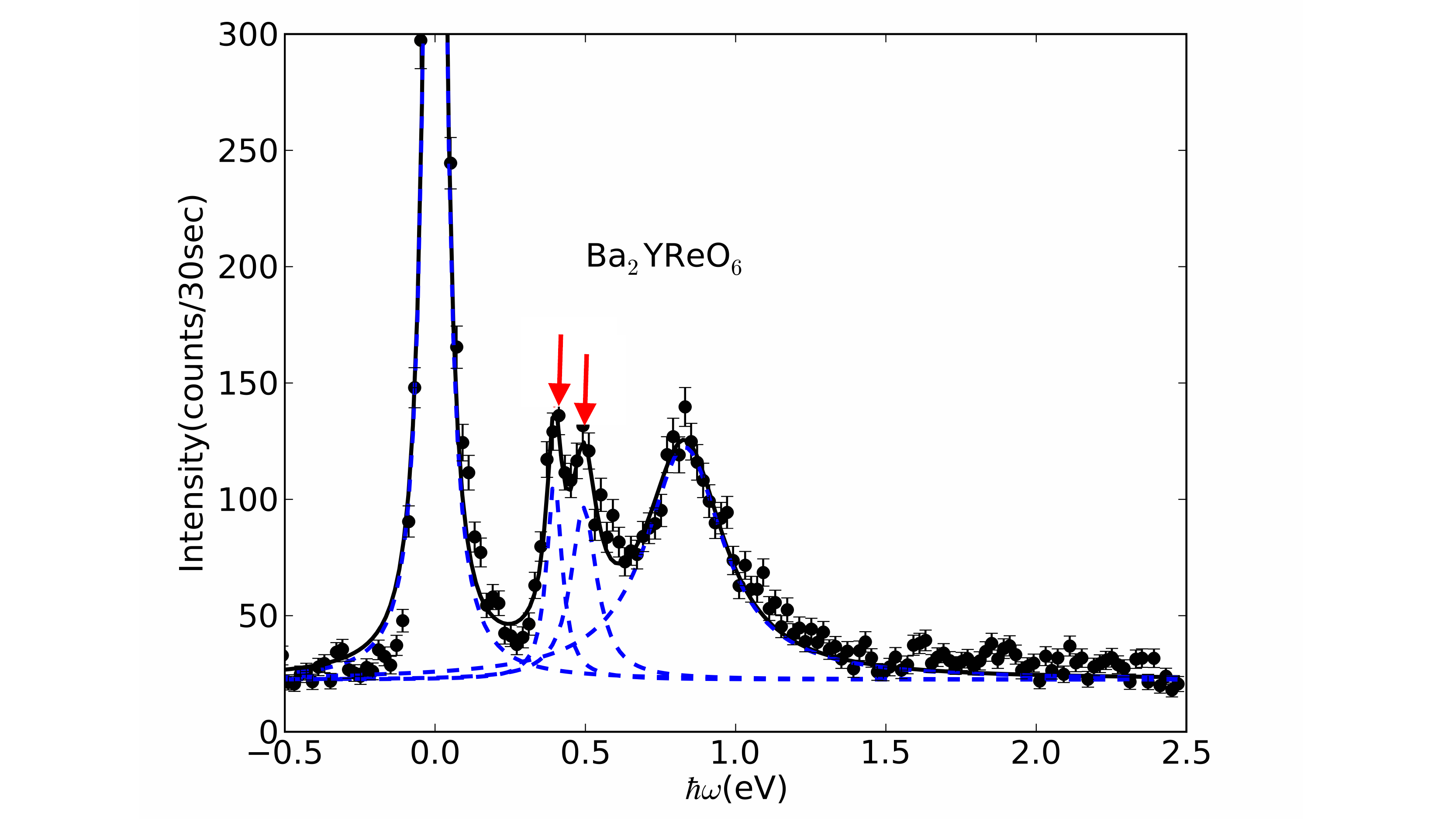}
\caption{(Color online) RIXS intensity as a function of energy transfer $\hbar\omega$ in Ba$_2$YReO$_6$. The RIXS spectrum was obtained near Re $L_3$ edge with  incident energy, E$_i=10.537$keV. A scattering geometry with $2\theta$=90$^{\circ}$ was used to minimize elastic background. The two indicated peaks at $\hbar\omega = 0.40$eV and $0.50$eV 
were unresolved in previous $L_2$ edge measurements.}
\label{newdata}
\end{center}
\end{figure}

\section{Ab initio electronic structure calculations}

We carried out density functional theory (DFT) calculations
for the cubic double perovskites Ba$_2$YReO$_6$, Ba$_2$YOsO$_6$, and Ba$_2$YIrO$_6$. 
From a structural viewpoint, these cubic
double perovskites can be regarded as consisting of
clusters of the metal atom ${\cal M}$ with the six O ions comprising the (${\cal M}$O$_6$)$^{7-}$ 
octahedral cage, separated by Ba and Y ions that maintain the charge balance.
We used the
generalized gradient approximation (GGA) of Perdew, Burke and Ernzerhof (PBE)
\cite{pbe}, with the addition of an on-site Coulomb repulsion using
the PBE+U method ($U$= 4 eV) in the so-called fully localized limit,
and the general potential linearized augmented planewave (LAPW) method
\cite{singh-book} as implemented in the WIEN2k code. \cite{wien2k}
In this method U is a parameter applied in order to mimic the effects of Coulomb correlations \cite{Anisimov1997}. 
This value is applied in the LAPW method to the d-orbitals within an LAPW sphere. Typical values for transition metal 
oxides range from 4 eV to 8 eV. In the present case, where we deal with a multi-orbital 5d material, low values are likely to 
be more physical. We find that 3 eV is inadequate to give an insulating gap for all the compounds studied, while 
experimental resistivity data suggests insulating character. We choose 4 eV because this is adequate to give an insulating gap 
in all compounds at least for an AFM state.
Further details of the DFT calculations are given in Appendix A.
Note that in our {\it ab initio} electronic structure calculations, we rely on experimental lattice parameters to fix atomic positions 
because they are well established for these materials and are without doubt more accurate than can be obtained 
from DFT.

For all three materials, we have studied $5d$ moments arranged in a type-I antiferromagnetic (AFM-I) pattern
and a ferromagnetic (FM) pattern. The calculated DOS for FM order in different compounds are given in the Appendix.
For FM order, the DOS peaks are generally broader, leading to incomplete gapping for the Ir and Re compounds.
We also considered non-spin-polarized solutions
which, however, are not energetically favored for any of the compounds studied even with
U = 0 eV. This argues against explanations for the lack of observed magnetic
ordering in the Ir and Re compounds that rely on the absence of moments.

\subsection{Crystal field splitting}

Fig. \ref{dos} shows the metal $d$ projection of the calculated density
of states (DOS) for all three compounds, including the spin projections, in the AFM-I state. 
We see that all compounds show a very large crystal field splitting, with $t_{2g}$ states, which are
near the Fermi energy ($E=0$), being well separated from the $e_g$-like states at $\approx \pm 5$eV. These
$e_g$-like states correspond to strongly hybridized bonding and anti-bonding combinations of 
$5d$ $e_g$ states and O $2p$ states arising from a significant $\sigma$-overlap. This large
crystal-field splitting is consistent with RIXS data.\cite{Taylor_PRL2017, Yuan_PRB2017,Nag2017}
Since the top of the $t_{2g}$ DOS is separated from the bottom of the $e_g$-like
DOS by $\approx 2$eV, a model based on just the $t_{2g}$ orbitals, as we have discussed above, 
is appropriate to understand the RIXS spectra for energy transfers $\hbar\omega \lesssim 2$eV
in all the compounds. However, as we discuss next, a detailed examination of the
spin and charge distribution within the (${\cal M}$O$_6$)$^{7-}$ cluster, and a study of different magnetic
ordering patterns, reveals interesting physics beyond the atomic limit.

%At the same time, we note that there
%is no evidence of strong direct hybridization of the 5$d$ $t_{2g}$ states (which are located near the Fermi energy) with
%the O $p$ states. For example, one sees little $5d$ character in the upper
%O $p$ bands ($\approx 2$-$3$eV below the Fermi energy), where $t_{2g}$-O$2p$ $\pi$-antibonding states would
%be expected. Thus, our DFT calculations provide
%clear justification for projecting the physics of the RIXS excitations 
%with $\hbar\omega \lesssim 2$eV to the t$_{2g}$ orbitals of the $5d$ ion as in our model calculations. 

\subsection{Spin and orbital moments}

For $U=4$eV, we find that all three compounds are insulating and show local moment behavior in the 
sense that the spin and orbital moments on the metal site are practically identical
for the AFM and FM orders. A summary of the moments is given in Table \ref{table-moments}.

We start with a discussion of the spin-moment. As seen from the values of $M_{spin}$, the total spin in the unit
cell in the FM pattern, SOC only weakly reduces the total
spin-moments from the nominal values of 2 $\mu_B$/atom for Re and Ir, and
3 $\mu_B$ for Os (based on electron count in an isolated $t_{2g}$ shell).
However, the moments as quantified by the part residing in the metal sphere, $m_{spin}$,
are only $\approx\! 2/3$ of the total moment, with the strongest reduction
(to $\sim\! 60\%$) for the Ir case. This deficit is because 
some of the moment in the (${\cal M}$O$_6$)$^{7-}$ cluster
is on the O site. (Within our DFT calculations for Ba$_2$YIrO$_6$, 
decreasing $U$ leads to moment reduction on the Ir site, which may bring it in closer alignment
with susceptibility measurements. However, we find that this also leads to a metallic
DOS, in apparent contradiction with transport data. We return to this issue later.)

\begin{figure}[tbh]
\includegraphics[width=0.48\columnwidth]{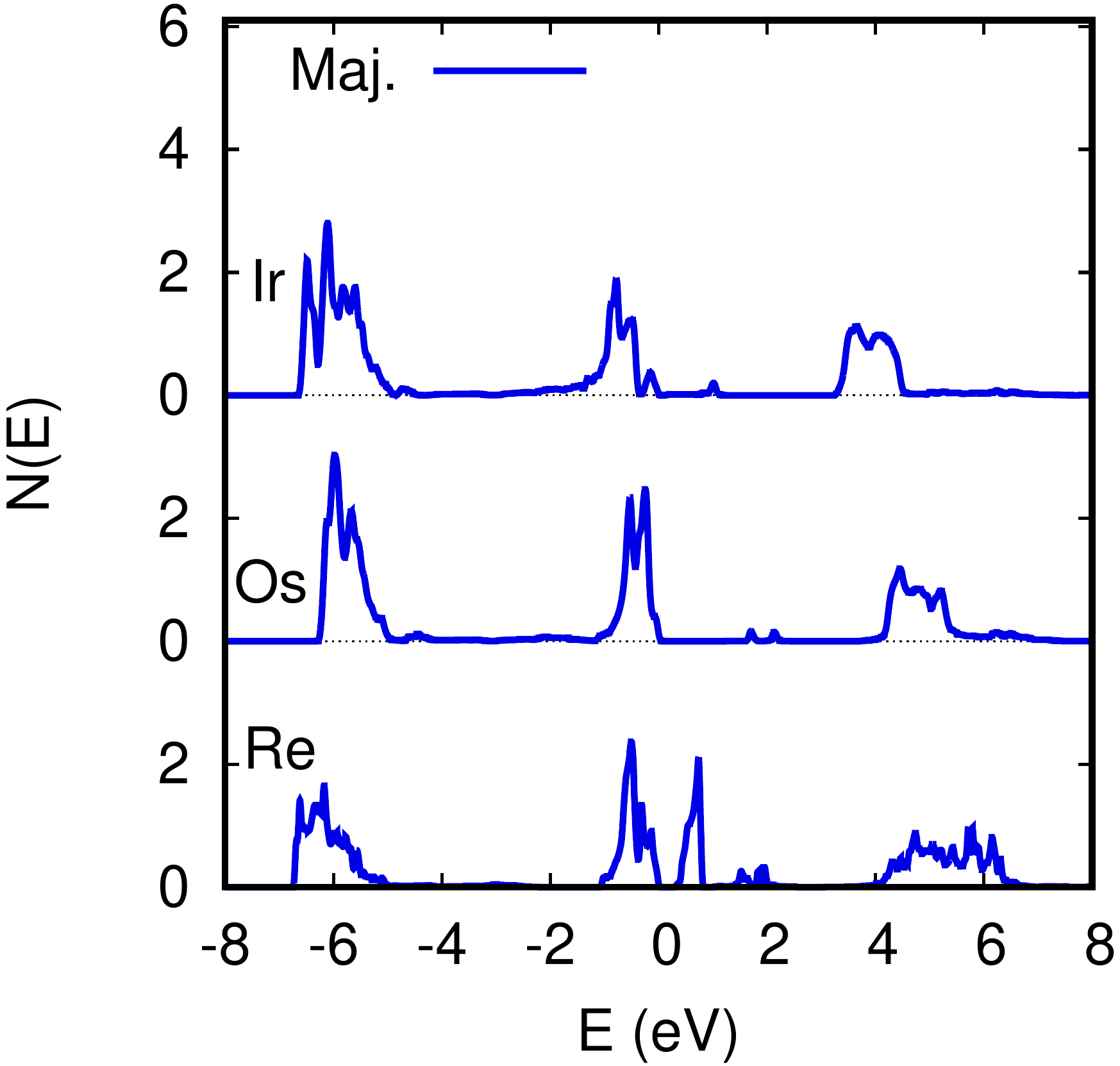}
\includegraphics[width=0.48\columnwidth]{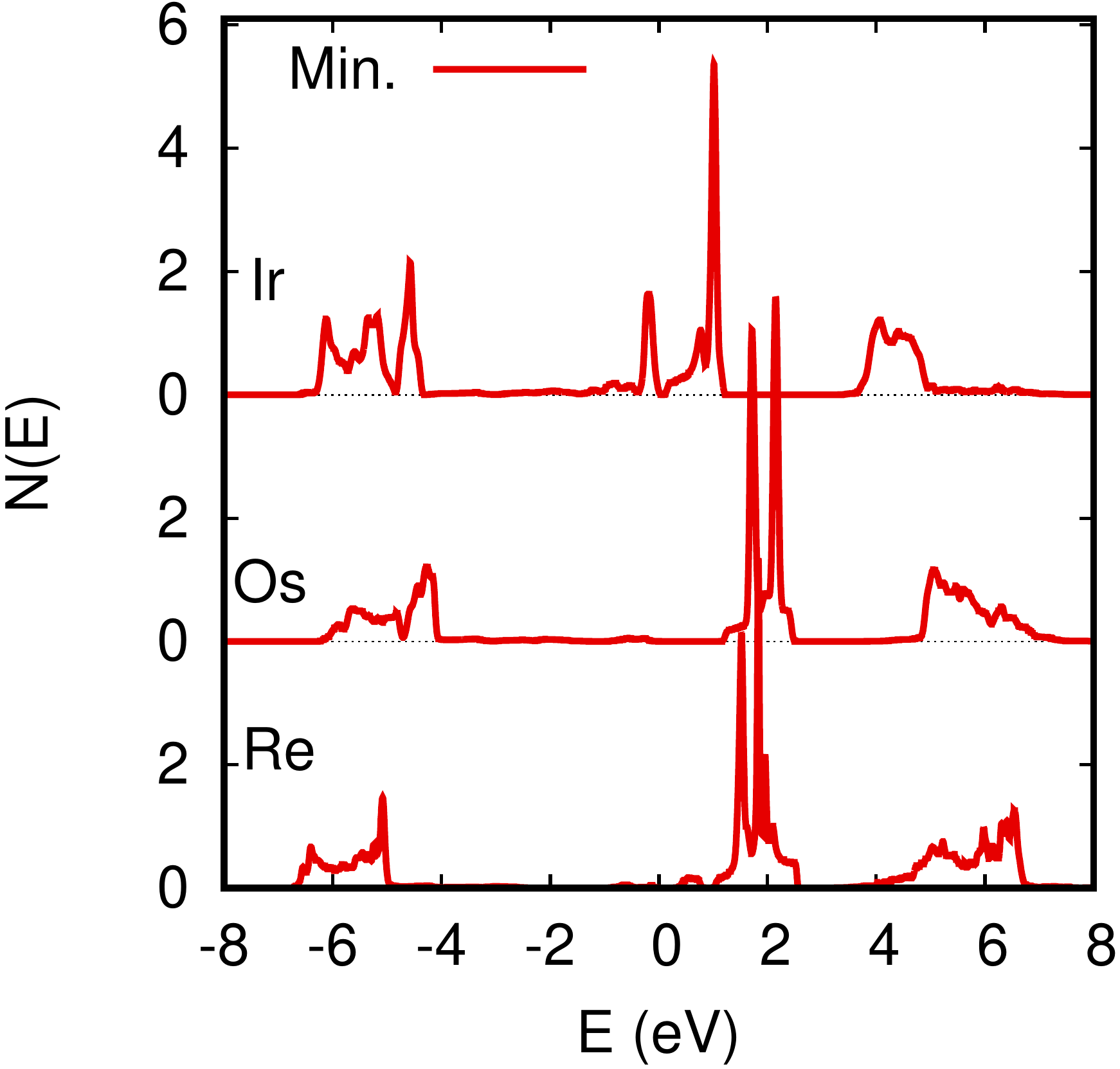}
\caption{(Color online) 5$d$ projections of the electronic density of states onto the
Ir, Os, and Re LAPW spheres of majority and minority spin character on a
per ion basis. The Os values are offset and the energy zero is at the
highest occupied state.}
\label{dos}
\end{figure}

It is also interesting to note that while the Re and Os
compounds have orbital moments in accord with the {\it ionic} Hund's rule (i.e.,
opposite to the spin moment), this is not the case for the Ir compound. This is a consequence of
the strong crystal field splitting noted above, between the $t_{2g}$ and $e_g$
orbitals.
This reversal of the orbital moment for the Ir compound does not follow the third Hund's rule for a free ion, 
but does follow the Hund's rule if one considers the t$_{2g}$ orbital as an independent shell (i.e. as an 
effective p level, which is now more than half full for the Ir compound).

Finally, we turn to the issue of why DFT finds large induced O moments
in these compounds. The explanation lies in an indirect exchange mechanism
where the on-site Hunds exchange coupling couples the $t_{2g}$ moments
to produce an exchange splitting of the $e_g$ 5$d$ orbitals,
which occur both in the $e_g$ upper crystal field level and at the bottom of
the O $2p$ bands as seen from Fig.~\ref{dos}. This leads to the spin dependent hybridization of the
$e_g$ orbitals, and therefore a magnetization of the nominally O $2p$ derived bands.

We note that the LAPW method divides space into non-overlapping spheres centered at the atoms and a remaining interstitial space. The O spheres in our calculation are necessarily small due to this non-overlapping requirement, and therefore the moment in these spheres underestimates the O contributions, but is expected to be roughly proportional to them. The contribution from the LAPW spheres of the six O around a given transition metal atom are $0.20 \mu_B$, $0.47 \mu_B$, and $0.44 \mu_B$, for the Re, Os and Ir compounds, respectively. Note that the total of the O and transition metal atoms is not the total moment due to the interstitial, and also that even with a gap, the spin-orbit interaction reduces the total spin moments from nominal integer values that may be expected from the band filling.

\begin{table}[tb]
\caption{Calculated spin and orbital moments $m$ (in $\mu_B$) in the transition
metal LAPW sphere for AFM and FM
arrangements (see text) from DFT calculations with $U$=4 eV.
$M_{spin}$ is the total
spin moment per formula unit including all atoms for the FM case.
$\Delta E = E_{FM}-E_{AFM}$ is the energy difference (per formula unit) between FM and AFM states.\\}
\begin{tabular}{|l|cc|ccc|c|}
\hline
~~Material  &  ~~~AFM & & & FM & & $\Delta E$ \\
  & $m_{spin}$ & $m_{orb}$ & $m_{spin}$ &$m_{orb}$ & $M_{spin}$ & (meV) \\
\hline
Ba$_2$YReO$_6$ & 1.27 & -0.59 & 1.30 & -0.66 & 1.97 & 3.6 \\
Ba$_2$YOsO$_6$ & 1.87 & -0.12 & 1.89 & -0.13 & 2.95 & 54.5 \\
Ba$_2$YIrO$_6$ & 1.09 & +0.40 & 1.09 & +0.39 & 1.87  & 0.4 \\
\hline
\end{tabular}
\label{table-moments}
\end{table}

\subsection{Magnetic ordering}

Our DFT calculations yield magnetic ground states in all three compounds,
in that the AFM-I structure gives lower energy than a non-magnetic case.
This result is robust against changes in the parameter $U$, and in particular
also holds for $U$=0. However, we find that the exchange interaction between (${\cal M}$O$_6$)$^{7-}$
clusters, as quantified by the AFM-FM energy difference, while always antiferromagnetic, is
one to two orders of magnitude smaller in the Re and Ir compounds as compared
to the Os compound. This may be important for explaining
experiments showing evidence
for the presence of moments in the Ir and Re compounds, but without the robust long range order
observed in the Os compound.

For Ba$_2$YOsO$_6$, the magnetic structure has been experimentally determined  \cite{kermarrec} to be type-I AFM order
below a N\'eel temperature $T_N = 69$K.
Indeed our results show that AFM-I order leads to a lower energy than FM order. Based on the 
energy difference $\Delta E = E_{FM}-E_{AFM} = 54.5$meV per Os, we infer a high Curie-Weiss temperature $\Theta_{CW} \gtrsim 600$K,
consistent with experiments.\cite{kermarrec}

For Ba$_2$YReO$_6$, we find a much smaller energy difference $\Delta E = 3.6$meV,
so that we expect magnetic ordering tendencies are much weaker.
Experimentally, Ba$_2$YReO$_6$ is reported to show a glassy magnetic ground state possibly without long
range order and without evidence in thermodynamics or susceptibility for a fluctuating state.\cite{sasaki,aharen,thompson}
We suppose that an AFM-I state may be the true ground state if a perfectly chemically ordered sample could be made,
with the observed glassy state resulting from low levels of disorder. 
Oxygen vacancies, if present in large quantities, might also provide a source of disorder affecting ordering.

The results on Ba$_2$YIrO$_6$ are still controversial,\cite{CaoPRL2014, Dey2016,Hammerath2017,GangCao2017,Nag2017,QChen2017,Kusch2018} with 
experimental reports of magnetism being attributed to impurities
or to weakly fluctuating $\sim\!0.4\mu_B$ moments whose origin is unclear. Previous electronic structure calculations and model studies\cite{CaoPRL2014,Bhowal2015,Trivedi2017,Hammerath2017,Gong2018} reach somewhat 
contradictory conclusions based on whether one starts from a band picture or an atomic picture.
From our calculations with $U=4$eV, we
find a significant moment $\sim\! 1.5\mu_B$ on the (IrO$_6$)$^{7-}$ cluster, but with $\Delta E = 0.4$meV 
which would imply a negligibly small exchange coupling between moments on neighboring clusters.
The value of the energy difference is sensitive to the parameter $U$, but we verified that it remains much smaller than in the Os 
compound for different values. At some values of $U$ (e.g. $U = 3$eV) the ferromagnetic order can even have lower energy than the AF-I order.
Assuming that the experimentally measured moments in Ba$_2$YIrO$_6$  are indeed intrinsic, our calculations could help to understand why these
moments may not order down to very low temperatures.

We note that the very small energy difference between ferromagnetic and antiferromagnetic orderings means that the inter-site exchange couplings are small, 
which is the reason for inferring weak magnetic interactions. As noted previously \cite{Mazin1997,Calder2017}, 
in 4d and 5d double perovskites, oxygen takes a substantial spin polarization 
leading to effective {\cal M}O$_6$ octahedral magnetic clusters. These interact through the O atoms so that the contact and distances between O in different octahedra is important for the exchange. This suggests a sensitivity to structure. It will be of interest to experimentally explore strain and pressure effects on magnetic order in these compounds especially to better understand the non-ordered states of the Re and Ir compounds; this is a topic for future investigation.

\section{Discussion}

Our ED results show that the RIXS excitations in all the $5d$ double perovskites are well described by the atomic limit picture. In this limit, the $d^2$ rhenates and 
$d^3$ osmates should support local moments, which is consistent with our complementary electronic structure calculations. In addition, our {\it ab initio} estimates 
for the exchange interaction strength is consistent with experiments which find robust magnetic order in Ba$_2$YOsO$_6$ as opposed to Ba$_2$YReO$_6$.
However, the ED and DFT calculations are in disagreement for the ground state of Ba$_2$YIrO$_6$. While the atomic limit ground state in ED is a $J_{\rm eff}=0$ singlet,
our DFT results indicate that the $d^4$ iridates should show a significant local moment in the insulating phase.

Within our DFT calculations on Ba$_2$YIrO$_6$, we find that decreasing $U$ leads to a smaller moment on the Ir site. 
This could partially bridge the gap with ED, and may bring the moment in closer alignment
to that inferred from susceptibility measurements,\cite{CaoPRL2014,Dey2016}  However, the resulting state then becomes metallic which
seems to be at odds with the apparently insulating resistivity,\cite{Hammerath2017} unless we ascribe
this to disorder induced localization. Assuming that the insulating transport is intrinsic and due to interactions, we are led to conclude that quantum 
spin-orbital fluctuations and dynamical self-energy effects beyond DFT must be crucial in Ba$_2$YIrO$_6$.
Including these may lead to one of two outcomes.
(i) This could stabilize a Mott insulator with small moments which are weakly coupled, which could explain both the susceptibility and transport data,
showing that going beyond the simple atomic limit is important.
(ii) Alternatively, it might stabilize the $J_{\rm eff}=0$ state as in our ED study; the measured magnetism must then
be attributed to defects.\cite{Hammerath2017,QChen2017}

\begin{figure}[t]
\includegraphics[width=\columnwidth]{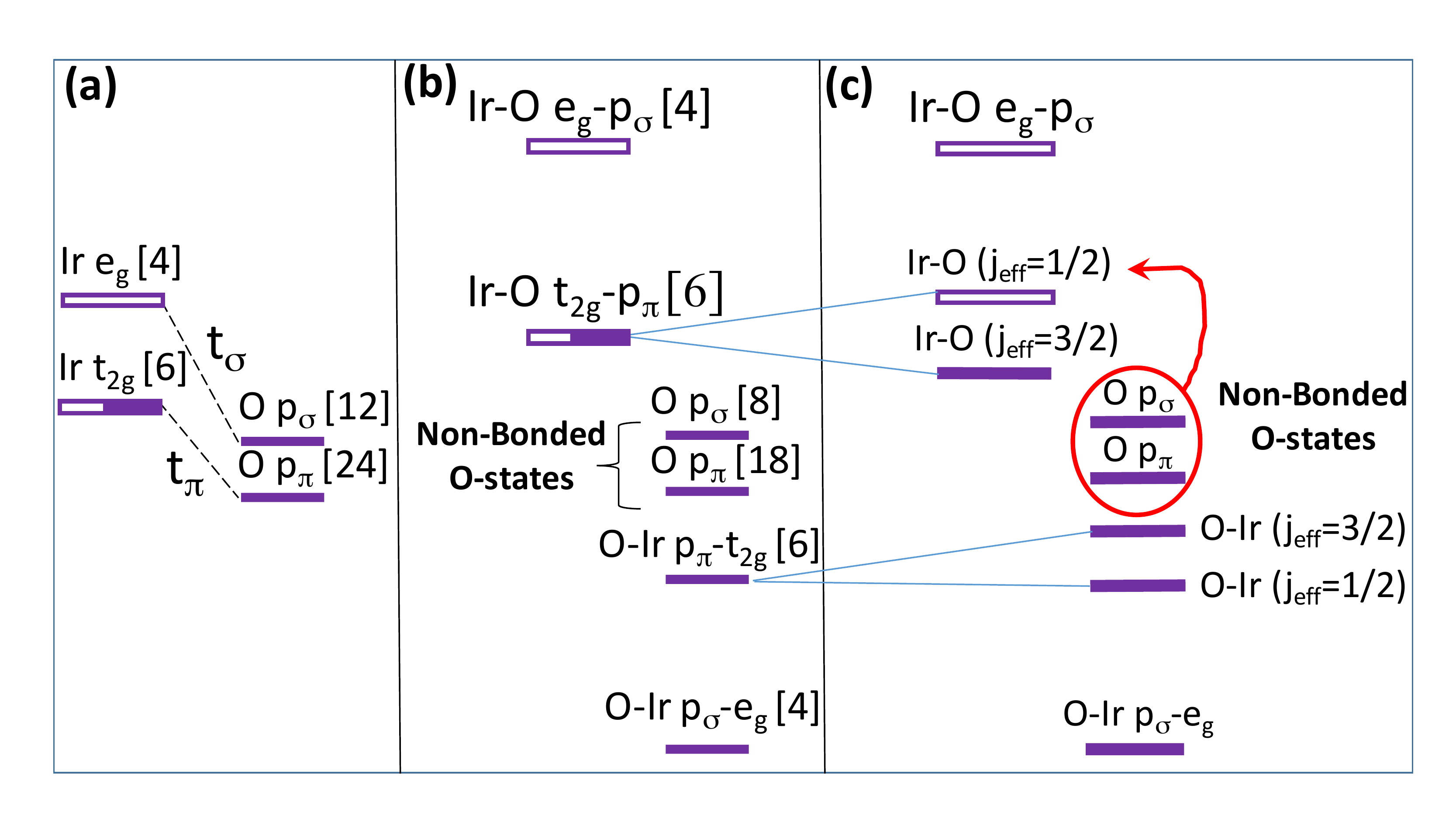}
\caption{(Color online) Schematic single-particle level diagram for Ir and O orbitals in the IrO$_6$ cluster of Ba$_2$YIrO$_6$, with
filled, partially filled, and empty boxes indicating electron filling.
(a) Ir and O orbitals showing small crystal field splitting in the absence of hybridization;
dashed black lines show intersite hybridization via hopping matrix elements $t_\sigma, t_\pi$. Numbers indicate degeneracies of the orbitals including spin. (b) Hybridization 
leads to a large splitting of the $\sigma$-hybridized orbitals, and to a smaller splitting of the $\pi$-hybridized orbitals, together with a set of unbonded O orbitals. The notation
Ir-O or O-Ir indicates respectively that the Ir or O states are the dominant contribution to the hybridized wavefunction. (c)
SOC (thin blue lines) on Ir leads to a splitting of the hybridized Ir-O and O-Ir $t_{2g}$ states. Nominally filled and unfilled levels are indicated by 
filled and empty boxes in this final schematic, which would lead to a $J_{\rm eff}=0$ insulator. Arrows shows possible impact of interactions which might cause
a partial occupancy of the $j_{\rm eff}=1/2$ hybridized  Ir-O level; such ``intracluster excitons'' may lead to a Mott insulator
with weak magnetic moments.}
\label{fig:Level}
\end{figure}

At the same time, in order to understand potentially how the atomic limit picture might weakly break down in Ba$_2$YIrO$_6$, it is useful to 
study an isolated IrO$_6$ octahedral cluster which allows for some degree of electron delocalization in the Mott insulating phase. Within a
perfect octahedral cage, the $e_g$ orbitals of Ir will each hybridize with one appropriate symmetry combination of the
$p_\sigma$ oxygen orbitals. Similarly, each $t_{2g}$ orbital can hybridize with only one symmetry combination of the $p_\pi$ orbitals. 
Figs.~\ref{fig:Level}(a),(b) present an illustrative level scheme where we have shown how the $\sigma$-hybridization leads to
Ir-O $e_g$ levels which are strongly split, while the smaller $\pi$-hybridization of the $t_{2g}$ states with a subset of O $p_\pi$,
leaving a residual set of non-bonded O levels. For hybridized states, we have used the notation Ir-O and O-Ir to respectively depict states
which are dominantly Ir versus dominantly O. Here, the numbers indicate the level 
degeneracy (including spin). Incorporating SOC, as shown in Fig.~\ref{fig:Level}(c), leads to hybridized $j_{\rm eff}=1/2,3/2$ states. Based on this 
final level scheme, all states upto and including
Ir-O $j_{\rm eff}=3/2$ are filled, while the Ir-O $j_{\rm eff}=1/2$ and antibonding  Ir-O $e_g$ states are unfilled, leading to a $J_{\rm eff}=0$ ground state. 
This level scheme is consistent with the `atomic' limit, 
with the effective SOC, as determined from the separation between the Ir-O $j_{\rm eff}=3/2$ and $1/2$ states, being set by the atomic SOC and the 
degree of Ir-O hybridization.

If Ba$_2$YIrO$_6$ is a Slater insulator with AFM order, electron itinerancy is expected to lead to a mixing of the $j_{\rm eff}=1/2,3/2$ levels in forming
bands. However, as discussed earlier, it is unclear if this picture holds, since the charge gap $\gtrsim 100$meV inferred from transport measurements is 
much larger than the tiny energy scale (few Kelvin) for
magnetism.\cite{Hammerath2017}
On the other hand, if Ba$_2$YIrO$_6$ is a Mott insulator, with the weak moments found in experiments indeed being intrinsic, 
it is clear that the tiny Ir-Ir superexchange, which is far smaller than SOC, cannot destabilize the $J_{\rm eff}=0$ singlet and lead to these
moments. So the `exciton condensation' mechanism for $J_{\rm eff}=0$ Mott insulators studied in Refs.~\onlinecite{Khaliullin2013,Trivedi2015} cannot be operative here.
However, the origin of the moments could arise from Ir-O interactions within an IrO$_6$ cluster.
In this case, our level scheme 
suggests some potential intrinsic mechanisms
to explain the magnetic moments reported in the Mott insulator. 
For instance, electron interactions might lead to a partial depletion of the non-bonded O levels just below the Fermi level and partial occupation 
of the $j_{\rm eff}=1/2$ Ir-O level. This is one way in which going beyond the atomic limit in the ED calculations might lead to a breakdown of the
strict $J_{\rm eff}=0$ picture for the Mott insulator, resulting in the formation of
weak moments. This state may be schematically represented as $|{\rm Ir}^{5+} \ra |{\rm O_{nb}~no\!-\!hole}\ra + 
\epsilon | {\rm Ir}^{4+} \ra |{\rm O_{nb}~1\!-\!hole}\ra$,
where `O$_{\rm nb}$' denotes non-bonded oxygen orbitals. Alternatively, interactions which generalize the rotationally invariant Kanamori form may lead to mixing of the form 
$|J_{\rm eff}=0 \ra + \epsilon |J_{\rm eff}=1\ra$ on the Ir site since the Ir-O wavefunctions are not strictly ionic but represent states hybridized with oxygen.
In analogy with previous work,\cite{Khaliullin2013, Trivedi2015} we may term such states as `intracluster excitons'. Such excitons would be dispersionless, 
with extremely weak coupling between
clusters leading to possible weak long-range magnetic order. Further studies of such a cluster Hamiltonian would be valuable in exploring this scenario, since
it is unclear if these intrinsic explanations for the observed moments can also simultaneously be as successful at describing the RIXS data as our present model.

\section{Summary}

We have shown that the theory of RIXS yields mode energies and spectral intensities for $5d$ complex oxides at
different fillings which are in good
agreement with experiments, leading to estimates of SOC and Hund's coupling.
Our work provides a natural interpretation of the low energy peaks as single-particle excitations across the spin-orbit
gap, which are split by Hund's interaction, and the higher energy peaks as emerging from two-particle excitations across the spin-orbit gap 
which also leads to a lower intensity.
We note that recent work on $4d$ and $5d$ oxides
suggests that $t_{2g}$-$e_g$ interactions might become important for certain parameter regimes,\cite{Fiete_2017} thus going beyond the
approximation of projecting to the $t_{2g}$ orbitals. Our {\it ab initio} calculations show that the $e_g$ states
might also enter the picture differently, via strong hybridization with ligand oxygens. Our electronic structure calculations allow us to extract 
the exchange interactions, from which we deduce that Ba$_2$YOsO$_6$ should show robust AFM ordering, but that Ba$_2$YReO$_6$ and
Ba$_2$YIrO$_6$ have very weak exchange interactions which would strongly suppress magnetic ordering.
Finally, our electronic structure and atomic ED calculations lead us to a model for the
${\cal M}$O$_6$ cluster which may suggest a distinct mechanism for generating intrinsic weak magnetic moments in Ba$_2$YIrO$_6$.

AP, BY, and YJ were supported by the Natural
Sciences and Engineering Research Council of Canada. ADC was supported by the U.S. DOE, Office of Science, Basic Energy Sciences, 
Materials Sciences and Engineering Division

%\bibliography{RIXS}


\appendix

\section{Details of electronic structure calculations}

The DFT calculations were carried out with the generalized gradient approximation (GGA) of Perdew, Burke and Ernzerhof (PBE)
\cite{pbe} and the general potential linearized augmented planewave (LAPW) method
\cite{singh-book} as implemented in the WIEN2k code. \cite{wien2k} 
The LAPW sphere radii were 2.1 bohr for Ir, Os, Re and Y, 2.5 bohr for Ba and
1.55 bohr for O. We used the standard LAPW basis set plus
local orbitals for the semicore states.
With the PBE GGA, including magnetism, we obtain a semiconducting gap
for Ba$_2$YOsO$_6$, reflecting the exchange split $t_{2g}$ crystal
field level of this $d^3$ system, but we do not obtain a gap in either
Ba$_2$YReO$_6$ or Ba$_2$YIrO$_6$,
even with magnetic order and spin orbit coupling.
Experimental data (e.g. specific heat) imply that Ba$_2$YReO$_6$
is non-metallic. Experimental data is less clear for Ba$_2$YIrO$_6$
but it is presumed to be non-metallic based on transport data.
Accordingly, we show electronic structures with the PBE+U method,
with the choice $U$=4 eV. This is sufficient to open a gap in both
Ba$_2$YReO$_6$ and Ba$_2$YIrO$_6$. We find that with $U$=3 eV, a gap
is opened in Ba$_2$YReO$_6$ but not Ba$_2$YIrO$_6$, with the assumed
magnetic ordering pattern.
For $U$=4 eV, and the assumed antiferromagnetic order we obtain gaps of
0.31 eV, 1.21 eV and 0.12 eV, for the Re, Os and Ir compounds, respectively.
We note that the selected value of U is higher than that used by Bhowal et al. \cite{Bhowal2015} in a 
prior study of Iridates, where U = 2 eV was employed. In our calculations we find that
neither Ba$_2$YIrO$_6$ nor Ba$_2$YReO$_6$ is insulating for U = 2 eV. 
From an experimental point of view it is not fully established whether Ba$_2$YIrO$_6$ is a true 
insulator, but resistivity data points to such a state.

For the structure we used the
experimentally determined lattice parameters,
$a$=8.3395 \AA, \cite{aharen} for Ba$_2$YReO$_6$ and
$a$=8.34383 \AA, for Ba$_2$YOsO$_6$, \cite{Taylor_PRL2017}
and $a$=8.3387 \AA, for Ba$_2$YIrO$_6$. \cite{Dey}
We relaxed the free internal
parameter associated with the O position using the PBE GGA.
Since bonding and moment formation are inter-related, we allowed the
formation of ferromagnet moments in these relaxations, i.e. the FM order.
The resulting structures have Ba at (0.25,0.25,0.25) and (0.75,0.75,0.75),
Y at (0,0.0), Re/Os at (0.5,0.5,0.5) and O at (0,0,$z_{\rm O}$) and
equivalent positions,
with $z_{\rm O}$=0.2642, 0.2639 and 0.2636 for the Re, Os and Ir compounds,
respectively.
We used this structure for calculating electronic and
magnetic properties as discussed below.
All calculations included spin-orbit coupling,
except for the structure relaxation.

The majority and minority DOS for the optimal AFM state have been presented in the main text. Below, we plot the corresponding DOS
for the FM state.
\begin{figure}[tbh]
\includegraphics[width=0.48\columnwidth]{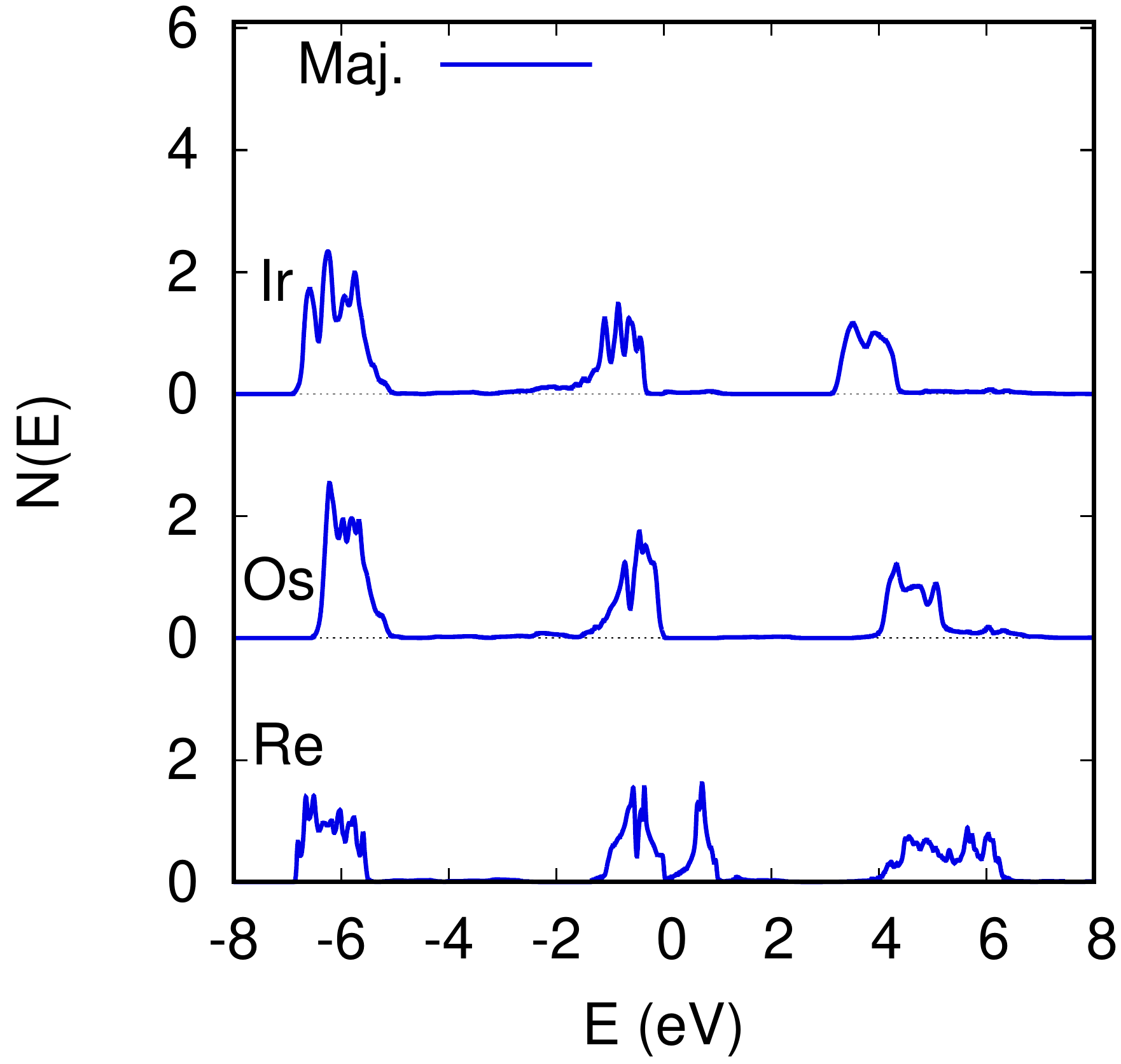}
\includegraphics[width=0.48\columnwidth]{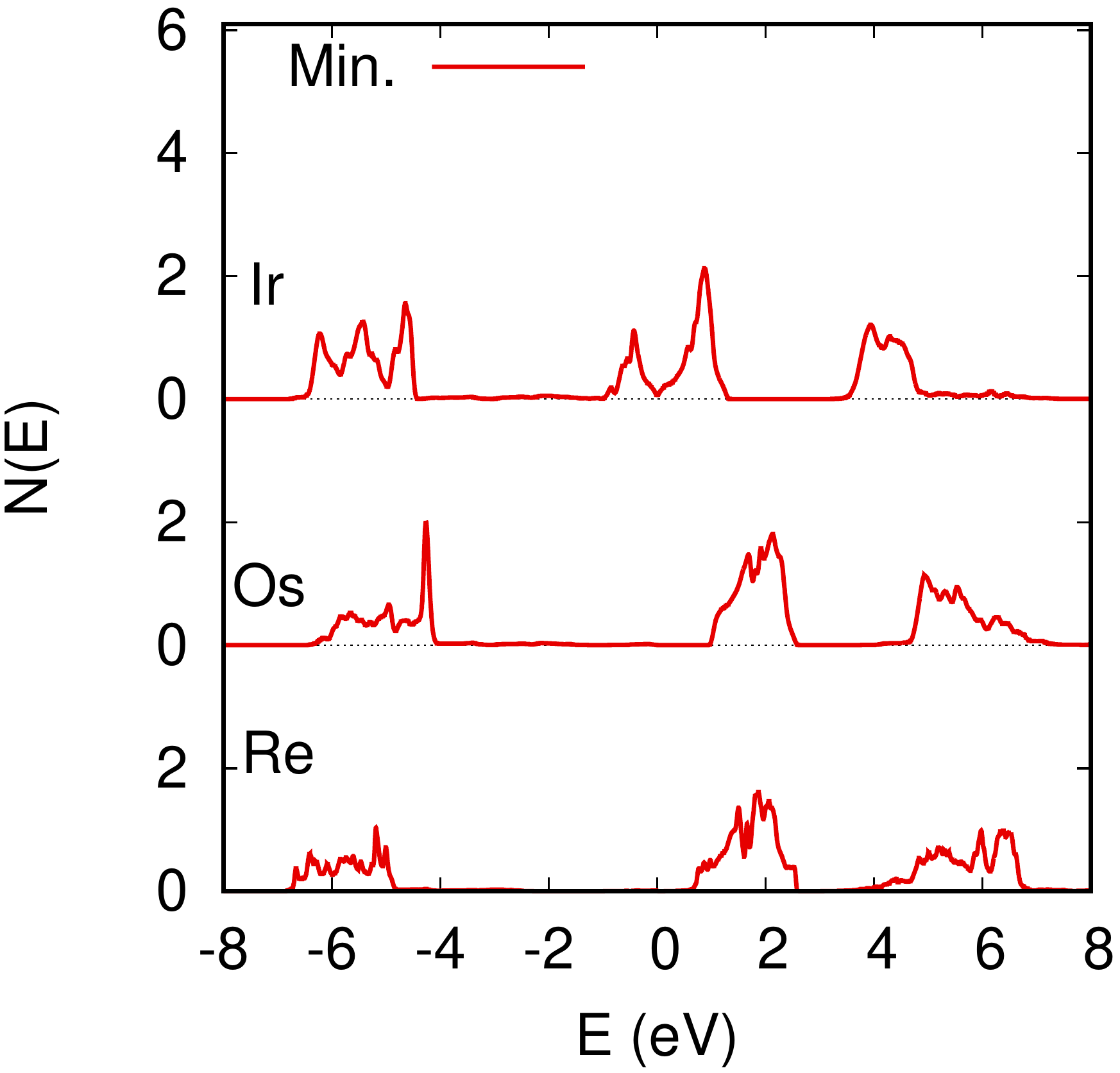}
\caption{(Color online) 5$d$ projections of the electronic density of states onto the
Ir, Os, and Re LAPW spheres of majority and minority spin character on a
per ion basis for the FM state. The Os values are offset and the energy zero is at the
highest occupied state.}
\label{dos}
\end{figure}

\end{document}